%% file: article.tex
\documentclass[12pt]{report}
\include{text/setups/setups}

\usepackage{caption}
\usepackage[english]{babel}
\usepackage{lineno, blindtext}
\usepackage{enumitem}
\usepackage{setspace}
\usepackage{float}
\floatstyle{plaintop}
\restylefloat{table}
\usepackage{bm}
\usepackage{booktabs}
\usepackage{array}
\usepackage{footnote}
\usepackage{afterpage}
\makesavenoteenv{tabular}
\makesavenoteenv{table}
\usepackage{footnotebackref}
\usepackage{amsmath}
\usepackage{colortbl}
\usepackage{makecell}
\usepackage{boldline}
\usepackage[toc,page]{appendix}
\usepackage{placeins}
\usepackage{chngcntr}
\newcolumntype{?}{!{\vrule width 1.5pt}}

\usepackage{multirow, hhline}
\usepackage{tablefootnote}

\usepackage{background}
\usepackage{lipsum}

\SetBgContents{CLAS12 Note 2019-003}
\SetBgScale{1.35}
\SetBgAngle{0}
\SetBgOpacity{1}
\SetBgColor{black}
\SetBgPosition{current page.south west}
\SetBgHshift{12cm}
\SetBgVshift{20cm}

\begin{document}

 \makeatletter
\newenvironment{sqcases}{%
  \matrix@check\sqcases\env@sqcases
}{%
  \endarray\right.%
}
\def\env@sqcases{%
  \let\@ifnextchar\new@ifnextchar
  \left\lbrack
  \def\arraystretch{1.2}%
  \array{@{}l@{\quad}l@{}}%
}
\makeatother

\newcommand\footnoteref[1]{\protected@xdef\@thefnmark{\ref{#1}}\@footnotemark}

\pdfcompresslevel=9
\pdfobjcompresslevel=9
\include{text/aliases/aliases}

\include{text/title_page/title}

\pagebreak


\include{text/toc/toc}

\include{text/introduction/intro}

\include{text/par_bosted/par_bosted}

\include{text/par_peak/par_peak}

\include{text/par_check/par_check}

\include{text/concl/concl}

\include{text/append/append}

\pagebreak
\newpage

\singlespacing
\bibliographystyle{ieeetr}
\bibliographystyle{apsrev4-1long}
\bibliography{note}

\end{document}

%% file: text/setups/setups.tex
\usepackage{rotating}
\usepackage[usenames]{color}
\usepackage{overpic}
\usepackage{url}
\usepackage{pst-plot}
\usepackage{amsmath,fullpage,graphicx,caption,float}
\usepackage{amssymb}
\usepackage{array,ragged2e,pst-node,pst-dbicons}
\usepackage{pstricks}
\usepackage{pst-node}
\usepackage{pst-blur}
\usepackage{comment}
\usepackage{color}
\usepackage{subfigure}
\usepackage{lscape}
\usepackage{multirow}
\usepackage{mathtools}
\usepackage{framed}
\usepackage{diagbox}

\usepackage{float}
\restylefloat{table}
\usepackage[margin=25mm]{geometry}

\renewcommand\baselinestretch{1.3}
\renewcommand\arraystretch{1.5}

\usepackage[numbers,sort&compress]{natbib}
\usepackage[nottoc]{tocbibind}

%% file: text/aliases/aliases.tex
 \def\be{\begin{eqnarray}}
 \def\ee{\end{eqnarray}}
 \def\ds{\displaystyle}

\newcommand{\mmm}{\mbox{íÜ÷}}
\newcommand{\reff}[1]{(\ref{#1})}
\newcommand{\ra}{\rangle}
\newcommand{\la}{\langle}
\newcommand{\rf}{\}}
\newcommand{\lf}{\{}
\newcommand{\ket}[1]{ | #1 \rangle }

%% file: text/title_page/title.tex
\noindent\begin{minipage}{\textwidth}
\begin{center}
\thispagestyle{empty}
\vspace{0.5cm}
{ \Large{Testing Parameterizations of the Deuteron Quasi-Elastic Peak}}\\
\vspace{1cm}

{\large Iu.A. Skorodumina$^{1, a}$, G.V. Fedotov$^{2}$,  R.W. Gothe$^{1}$} \\[16pt]

\parbox{.86\textwidth}{\centering\footnotesize\it
$^1$Department of Physics and Astronomy, University of South Carolina, Columbia, SC\\[8pt]
\setstretch{0.3} 
$^2$National Research Centre ``Kurchatov Institute" B. P. Konstantinov Petersburg Nuclear Physics Institute, Gatchina, St. Petersburg, Russia\\
[20pt]
E-mail: $^a$skorodum@jlab.org}\\

\vspace{2cm}
{\bf Abstract}\\[9pt]

\end{center}
{\small This study introduces common parameterizations of the quasi-elastic peak in the electron scattering spectrum off deuterium and provides the comparison of the parameterized cross sections with published experimental data. The comparison is performed in the wide $Q^{2}$ range from $\sim$0.3~GeV$^{2}$ to $\sim$4~GeV$^{2}$. In this way the performance of the parameterizations and their ability to describe experimental measurements are impartially tested and the conclusion on the description reliability is made.}

\end{minipage}

%% file: text/toc/toc.tex
\newpage
\renewcommand{\baselinestretch}{1}\normalsize
\tableofcontents
\setcounter{page}{2}

%% file: text/introduction/intro.tex
\newpage
\chapter{Introduction}
\mbox{}\vspace{-\baselineskip}

The elastic cross section is one of the most universal and well-understood characteristics of a scattering process. This observable, depending only on a few parameters, is easily measurable experimentally and can be relatively easy parameterized, and therefore represents a rather popular quantity of scientific interest. Being an important observable itself, the elastic cross section also often serves as a reference point for various complicated analyses of exclusive reactions with multiparticle final states, where this observable is extracted as an auxiliary quantity in order to verify both the correct normalization of the main result and the quality of the electron selection.

Nowadays the elastic cross section of electron scattering off a free proton is well-known\footnote[1]{See App.~\ref{app_formalism} for some details on the $ep$ scattering formalism.} over a wide kinematic range, since it has been intensively studied experimentally for decades and eventually almost perfectly described by parameterizations as that of Peter Bosted. This parameterization is based on the study from Ref.~\cite{Bosted:1994tm}, which provides an empirical fit to the world data for the proton elastic electromagnetic form factors in the range 0~GeV$^2<Q^2<$ 30~GeV$^2$ and to the neutron electromagnetic form factors in the range 0~GeV$^2<Q^2<$ 10~GeV$^2$. The ability of the Bosted parameterization to describe available experimental data on the elastic $ep$ cross section is demonstrated in Tab.~\ref{tab:epelast} of App.~\ref{app_epelas}. This table provides the comparison between measured cross section values taken from Refs.~\cite{Goitein:1970pz,Sill:1992qw,Christy:2004rc} and the corresponding parameterized values. The comparison reveals an excellent agreement between the experimental measurements and the parameterization within a few percent, which has a tendency to slightly worsen as $Q^{2}$ grows from $\sim$0.3~GeV$^{2}$ to $\sim$5~GeV$^{2}$.

It is also noteworthy that in several analyses of exclusive reactions off the free proton the comparison of the auxiliary extracted elastic cross sections with the Bosted parameterization was performed in order to check the overall normalization of analyzed observables as well as the quality of the electron selection~\cite{Fed_an_note:2007,Fedotov:2008aa,Fed_an_note:2017,Fedotov:2018oan,Isupov_note,Isupov:2017lnd}. Here, the study~\cite{Fed_an_note:2007,Fedotov:2008aa} performed for 0.2~GeV$^2<Q^{2}<$ 0.6~GeV$^2$ observed agreement between experimental and parameterized values within better than 5\%, the study~\cite{Fed_an_note:2017, Fedotov:2018oan} performed for 0.4~GeV$^2<Q^{2}<$ 1.0~GeV$^2$ observed $\sim$3\% agreement, while the study~\cite{Isupov_note,Isupov:2017lnd} performed for 2~GeV$^2<Q^{2}<$ 5~GeV$^2$ observed $\sim$10\% agreement.

Meanwhile, for electron scattering conducted off a nucleus the corresponding quantity of interest is the quasi-elastic cross section off nucleons. In contrast with the elastic spectrum off the free proton, which is discrete for a given beam energy and at fixed polar scattering angle, the quasi-elastic cross section off nucleons is continuously spread over the energy of the scattered electron. This smearing, caused by the motion of nucleons within a nucleus, forms a so-called quasi-elastic peak in the scattering spectrum. The position and shape of this peak contain information about the internal structure of nuclei.

Compared to elastic scattering off free protons, quasi-elastic scattering off nucleons in nuclei is less understood and lacking the same quality of theoretical description. Nonetheless, several techniques have been developed on this matter with the deuteron (as the lightest nucleus) being the most investigated target. 

This note introduces several existing parameterizations for the deuteron quasi-elastic peak and provides the comparison of the parameterized cross sections with published experimental data. The comparison is performed in the wide $Q^{2}$ range from $\sim$0.3~GeV$^{2}$ to $\sim$4~GeV$^{2}$. In this way the performance of the studied parameterizations and their ability to describe experimental measurements are impartially tested and the conclusion on the description reliability is made.

This examination, already interesting by itself, can be of great use for those deuteron target analyses that use the auxiliary extracted quasi-elastic cross section as a reference point in order to verify both the correct normalization of the main result and the quality of the electron selection.

%% file: text/par_bosted/par_bosted.tex
\newpage
\chapter{Peter Bosted Parameterization}
\mbox{}\vspace{-\baselineskip}

The set of Bosted parameterizations of inclusive cross sections off different targets~\cite{Bosted_fit,Bosted:2007xd} contains the modeling of the quasi-elastic cross section off various nuclei (including the deuteron). This modeling is based on the method of the quasi-elastic cross section estimation that was developed in the framework of the Relativistic Fermi Gas model and described in Refs.~\cite{Amaro:2004bs,Alberico:1988bv,Donnelly:1998xg}. The general idea of this method is sketched below. 

The differential cross section of quasi-elastic scattering of a nucleus can be calculated in the laboratory frame as\footnote[2]{See also App.~\ref{app_formalism} for the $ep$ scattering formalism.}
\begin{equation}
\frac{d^{2}\sigma}{d\Omega dE'} = F^{QE}\left [ \frac{d\sigma}{d\Omega} \right ]^{*}_{Mott}\left (v_{L}G_{L}^{QE} + v_{T}G_{T}^{QE} \right ),
\end{equation}
where 
\begin{itemize}
\item the quantity in the square brackets is the Mott cross section of the electron scattering off a point-like charge that is defined as
\begin{equation}
\left [ \frac{d\sigma}{d\Omega} \right ]^{*}_{Mott} = \left [ \frac{2\alpha E'}{Q^{2}}\cos \frac{\theta_{e'}}{2} \right ]^{2},
\label{eq:mott}
\end{equation}
with $E'$ and $\theta_{e'}$ being the energy and the polar angle of the scattered electron in the Lab frame, $Q^{2}$ the photon virtuality, and $\alpha=1/137$ the fine structure constant;   
\item the functions $G_{L}^{QE}$ and $G_{T}^{QE}$ are defined as
\begin{equation}
\begin{aligned}
&G_{L}^{QE}&=~&\frac{\kappa}{2\tau}\left ( ZG_{E_{p}}^{2}+NG_{E_{n}}^{2} \right )~~\textrm{and}\\
&G_{T}^{QE}&=~&\frac{\tau}{\kappa}\left ( ZG_{M_{p}}^{2}+NG_{M_{n}}^{2} \right ),
\label{eq:eee}
\end{aligned}
\end{equation}
where $\tau = \frac{|Q^{2}|}{4m_{N}^{2}}$ and $\kappa = \frac{q}{2m_{N}}$ with the photon momentum magnitude $q$ and the nucleon mass $m_{N}$. The quantities $Z$ and $N$ are the numbers of protons and neutrons in the nucleus, respectively, and $G_{E}$ and $G_{M}$ are so-called Sachs electric and magnetic form factors that are related to the charge and magnetization density of the corresponding nucleon, respectively;  
\item  $v_{L} = \left [\frac{\tau}{\kappa^{2}} \right ]^{2} $ and $v_{T} = \frac{\tau}{2\kappa^{2}} +\tan^{2}{\frac{\theta_{e'}}{2}}$ are the kinematic factors and
\item $F^{QE}$ is the nuclear scaling function.
\end{itemize}

In the Bosted parameterization~\cite{Bosted_fit,Bosted:2007xd} the Sachs form factors are calculated according to Ref.~\cite{Bosted:1994tm}, which provides an empirical fit to the world data for the proton elastic electromagnetic form factors in the range 0 GeV$^{2}$$< Q^{2} <$ 30 GeV$^{2}$ and to the neutron electromagnetic form factors in the range 0 GeV$^{2}$$< Q^{2} <$ 10 GeV$^{2}$.

For the case of scattering of a deuterium nucleus the Bosted parameterization~\cite{Bosted_fit,Bosted:2007xd} in its default implementation estimates the nuclear scaling function using a PWIA calculation and the Paris deuteron wave function (see Ref.~\cite{Bosted:2007xd} for details). For heavier nuclei it uses the following parameterization of the scaling function taken from Ref.~\cite{Bodek:2014pka},

\begin{equation}
F^{QE}(\psi') = \frac{1.5576}{K_{F}[1 + 1.7720^{2}(\psi' + 0.3014)^{2}](1 + e^{-2.4291\psi'})}.\label{eq:fqe_scaling}
\end{equation}

Here $\psi'$ is the scaling variable defined in Refs.~\cite{Bodek:2014pka,Amaro:2004bs} and $K_{F}$ is the nucleus Fermi momentum.

In general, the parameterization of the nuclear scaling function given by Eq.~\eqref{eq:fqe_scaling} is applicable for all nuclei from deuterium to lead~\cite{Bodek:2014pka}. In the Bosted parameterization for the case of a deuterium nucleus one can switch from the default way of the scaling function calculation to this alternative way upon minor modifications of the source code.

%% file: text/par_peak/par_peak.tex
\newpage
\chapter{Approximations of the peak cross section value}
\mbox{}\vspace{-\baselineskip}

The first consistent description of inelastic electron-deuteron scattering, which in contrast to earlier calculations took into account all important contributions to the electron-nucleon interaction as well as to the final state interactions between the outgoing nucleons, was Durand's theory~\cite{Durand:1959zz,Durand:1961zz}. In 1961 in Ref.~\cite{Durand:1961zz} Durand gave a simple approximation formula for the cross section at the quasi-elastic peak, which then was being widely used by experimentalists. This formula is given by
\begin{equation}
\left [ \frac{d^{2}\sigma}{d\Omega dE'} \right ]_{peak} = \left [ \frac{d\sigma}{d\Omega} \right ]_{Mott}^{*} ( 4.57\cdot 10^{-3}) \frac{m_{N}^{2}}{pE} \left (G_{E_{p}}^{2} + G_{E_{n}}^{2} + \frac{\tau}{\epsilon}\left [G_{M_{p}}^{2} + G_{M_{n}}^{2}\right ]  \right )\frac{1}{1+\tau},\label{eq:durand}
\end{equation}
where $\epsilon = \left (1 +2(1+\tau)\tan^{2}\frac{\theta_{e'}}{2} \right )^{-1}$, $p = q/2$ with the photon laboratory momentum magnitude $q$, $E=\sqrt{p^{2} + m_{N}^{2}}$, and all other quantities are defined as above\footnote[3]{The quantity $p$ here corresponds to the momentum of the final proton in the cms of outgoing nucleons. See discussions in Refs.~\cite{Durand:1959zz,Durand:1961zz,Budnitz:1969dt} on the validity of the approximation $p = q/2$ at the quasi-elastic peak.}. The numerical coefficient is in MeV$^{-1}$.


Since Durand's theory a lot of papers on inelastic electron-deuteron scattering~\cite{Kocevar:1967,Budnitz:1969dt,Hanson:1973vf} have tried to modify it with respect to one or the other point to get a still better understanding of the existing experimental data. Among them a very interesting is the study~\cite{Kocevar:1967}, which estimates some higher order contributions that stem from the use of complete relativistic kinematics and provides a different formula for the peak cross section (see Eq.~(50) in Ref.~\cite{Kocevar:1967})\footnote[4]{When using the peak cross section formulae in this study, the nucleon electromagnetic form factors $G_{E}$ and $G_{M}$ are estimated according to Ref.~\cite{Bosted:1994tm}.}.

%% file: text/par_check/par_check.tex
\newpage
\chapter{Testing the parameterizations with existing data}
\mbox{}\vspace{-\baselineskip}

The Bosted parameterization~\cite{Bosted_fit,Bosted:2007xd} and the approximations of the peak cross section~\cite{Durand:1961zz,Kocevar:1967} were tested on two sets of published data, i.e. the first set consists of six measurements obtained for beam energies from 0.5 to 1.6~GeV and $Q^{2}$ from 0.3 to 1.8~GeV$^2$~\cite{Hanson:1973vf}, while the second set includes two measurements obtained for beam energies of 9.8 and 12.6~GeV and $Q^{2}$ of 2.5 and 4~GeV$^2$, respectively~\cite{Rock:1991jy,Rock_SLAC}\footnote[5]{Actually, the second set~\cite{Rock:1991jy,Rock_SLAC} contains three more measurements at $Q^{2} = $ 6, 8, and 10~GeV$^2$, but they are not considered here because the quasi-elastic peak vanishes for such high $Q^2$ values.}. 

\begin{figure}[htp]
\begin{center}
\includegraphics[width=\textwidth]{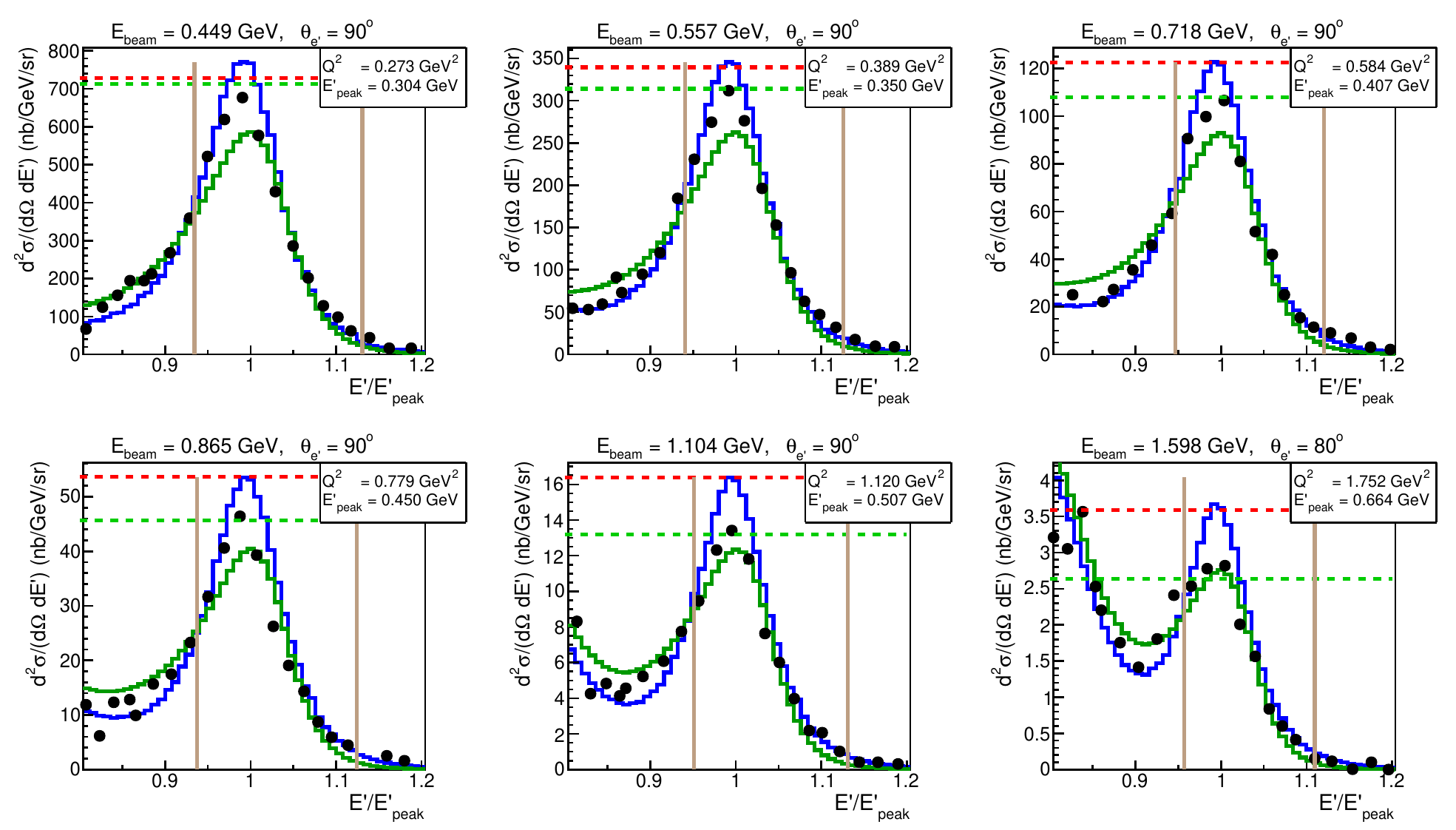}
\caption{\small  Data from Ref.~\cite{Hanson:1973vf} (black points) are compared with the Bosted parameterization~\cite{Bosted_fit,Bosted:2007xd} (histograms). The data uncertainties, which are on a level of 5\%, are not shown, see Ref.~\cite{Hanson:1973vf} on this matter. The blue histograms correspond to the default way of $F^{QE}$ calculation (using the Paris wave function), while the green ones correspond to $F^{QE}$ calculated by the alternative way (according to Eq.~\eqref{eq:fqe_scaling}). The green horizontal lines show the peak cross section values approximated by Eq.~\eqref{eq:durand}, while the red lines correspond to the predictions of Eq.~(50) from Ref.~\cite{Kocevar:1967}. The vertical lines show the integration limits.} \label{fig:hanson_QE}
\end{center}
\end{figure}

Figure~\ref{fig:hanson_QE} shows the quasi-elastic peak for the first set of measurements~\cite{Hanson:1973vf} (black points) compared with the Bosted parameterization shown as histograms. The blue histograms correspond to the default way of $F^{QE}$ calculation (using the Paris wave function), while the green histograms correspond to $F^{QE}$ calculated by the alternative way (according to Eq.~\eqref{eq:fqe_scaling}). As is seen from Fig.~\ref{fig:hanson_QE}, although describing rather nicely the left and right distribution slopes, the blue histograms systematically overestimate the peak values. Meanwhile, the green histograms have worse description of the left slope and have a tendency to underestimate the peak cross sections. The parameterization histograms were produced together with the inelastic part of the spectrum to facilitate visual comparison with the experimental measurements and to account for inelastic background under the quasi-elastic peak. The green horizontal lines in Fig.~\ref{fig:hanson_QE} correspond to the peak cross section values approximated by Eq.~\eqref{eq:durand}, while the red lines correspond to the predictions of Eq.~(50) from Ref.~\cite{Kocevar:1967}. As is seen, the former describe nicely the experimental peak values, while the latter match well the maxima of the blue histograms of the Bosted parameterization.

\afterpage{\clearpage}
\begin{figure}[htp]
\begin{center}
\includegraphics[width=14cm]{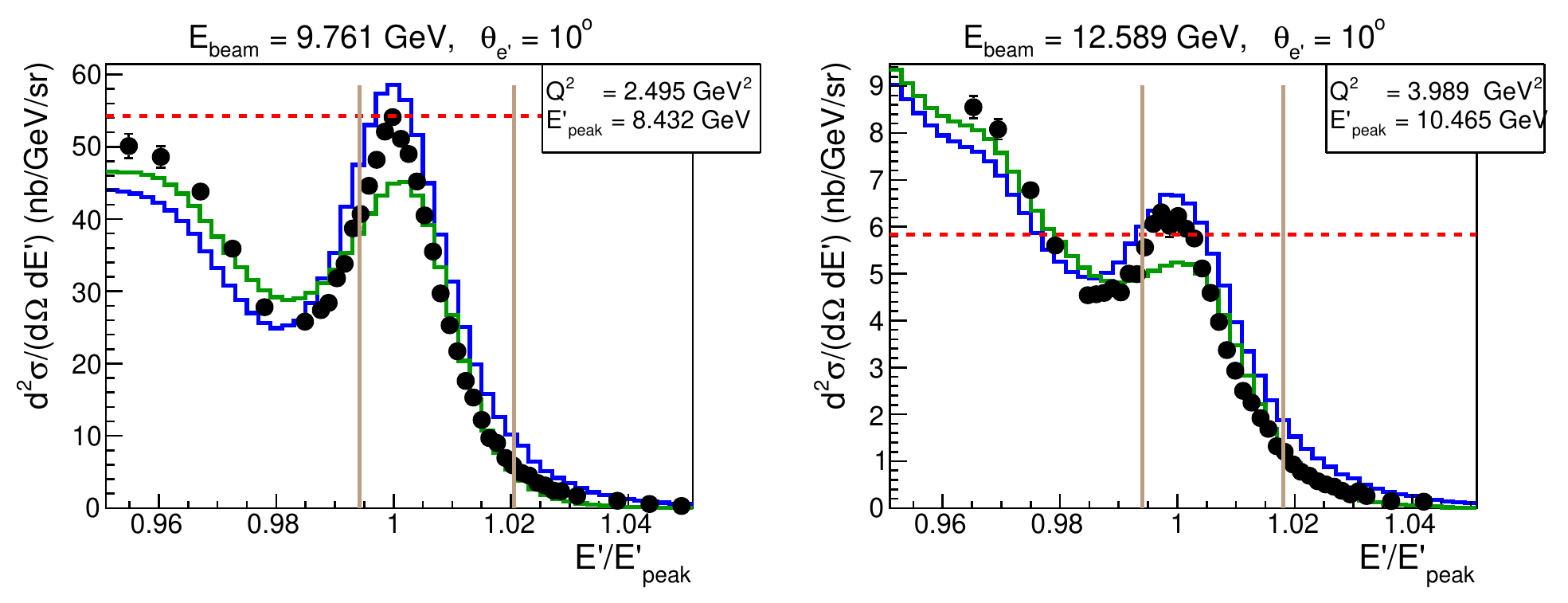}
\caption{\small Data from Refs.~\cite{Rock:1991jy,Rock_SLAC} (black points) are compared with the Bosted parameterization~\cite{Bosted_fit,Bosted:2007xd} (histograms). The blue histograms correspond to the default way of $F^{QE}$ calculation (using the Paris wave function), while the green histograms correspond to $F^{QE}$ calculated by the alternative way (according to Eq.~\eqref{eq:fqe_scaling}). The red horizontal lines show the peak values predicted by Eq.~(50) from Ref.~\cite{Kocevar:1967}. The vertical lines show the integration limits.  } \label{fig:rock_QE}
\end{center}
\end{figure}
\begin{table}[htp]
\begin{center}
\caption{\small Ratios of the experimental integrals under the quasi-elastic peak ($\sigma_{exp}$) to those obtained from the Bosted parameterization~\cite{Bosted_fit,Bosted:2007xd}, in which $F^{QE}$ is calculated by using the Paris wave function ($\sigma_{par}^{1}$) or given by Eq.~\eqref{eq:fqe_scaling} ($\sigma_{par}^{2}$). The first six rows correspond to the first dataset from Ref.~\cite{Hanson:1973vf} and the last two to the second dataset from  Refs.~\cite{Rock:1991jy,Rock_SLAC}. The index $norm$ means that the parameterization histograms were scaled in a way that their maxima were equal to the predictions of Eq.~\eqref{eq:durand} for the first dataset and to the predictions of Eq.~(50) from Ref.~\cite{Kocevar:1967} for the second dataset. The coloring of the table cells is related to the corresponding deviation of the obtained ratio from unity: the dark-green shade stands for deviations $\leq 5$\%, light-green for 5\%-10\%, and light-red for more than 10\%.  \label{tab:quasi_el_tab}}
\begin{tabular}{
   !{\vrule width 2pt}
  c!{\vrule width 1pt}
  c!{\vrule width 1pt}
  c!{\vrule width 1pt}
  c!{\vrule width 2pt}
  c!{\vrule width 1pt}
  c!{\vrule width 2pt}
  c!{\vrule width 1pt}
  c!{\vrule width 2pt}
  }
\toprule[2pt]
\makecell{Exp.\\ Ref.} &\makecell{$E_{beam}$\\ (GeV)} &\makecell{$Q^{2}$\\ (GeV$^2$)} & \makecell{$E'_{peak}$ \\(GeV)} &$\sigma_{exp}/\sigma_{par}^{1}$ &$\sigma_{exp}/\sigma_{\substack{par, \\ norm}}^{1}$& $\sigma_{exp}/\sigma_{par}^{2}$ &$\sigma_{exp}/\sigma_{\substack{par, \\ norm}}^{2}$ \\ \hline
\multirow{6}{*}{\cite{Hanson:1973vf}}&0.449  &0.273  &0.304  &\cellcolor{green!20}0.94   &\cellcolor{green!35}1.02   &\cellcolor{red!20}1.14    &\cellcolor{green!20}0.93\\ \hhline{|~|-------|}
&0.557  &0.389  &0.350  &\cellcolor{green!20}0.93   &\cellcolor{green!35}1.03     &\cellcolor{red!20}1.14    &\cellcolor{green!35}0.95\\ \hhline{|~|-------|}
&0.718  &0.584  &0.407  &\cellcolor{green!20}0.91     &\cellcolor{green!35}1.03   &\cellcolor{red!20}1.12  &\cellcolor{green!35}0.96\\ \hhline{|~|-------|}
&0.865  &0.779  &0.450  &\cellcolor{red!20}0.84     &\cellcolor{green!35}0.99   &\cellcolor{green!35}1.04  &\cellcolor{green!20}0.92\\ \hhline{|~|-------|}
&1.104  &1.120  &0.507  &\cellcolor{red!20}0.85     &\cellcolor{green!35}1.05   &\cellcolor{green!20}1.06  &\cellcolor{green!35}1.00\\ \hhline{|~|-------|}
\multirow{5}{*}{\cite{Rock:1991jy,Rock_SLAC}}&1.598  &1.752  &0.664  &\cellcolor{red!20}0.85     &\cellcolor{red!20}1.18   &\cellcolor{green!20}1.06  &\cellcolor{red!20}1.11\\ \midrule[2pt]
&9.761  &2.495  &8.432  &\cellcolor{red!20}0.83     &\cellcolor{green!20}0.90     &\cellcolor{green!35}1.05  &\cellcolor{red!20}0.87\\ \hhline{|~|-------|}
&12.589 &3.989  &10.465 &\cellcolor{red!20}0.84     &\cellcolor{green!35}0.97     &\cellcolor{green!35}1.04  &\cellcolor{green!20}0.93\\ \bottomrule[2pt]
\end{tabular}
\end{center}
\end{table}

Figure~\ref{fig:rock_QE} shows the quasi-elastic peak for the second set of measurements~\cite{Rock:1991jy,Rock_SLAC} (black points) compared with the Bosted parameterization shown as histograms. Again, the blue histograms correspond to the default way of $F^{QE}$ calculation (using the Paris wave function), while the green ones correspond to $F^{QE}$ calculated by the alternative way (according to Eq.~\eqref{eq:fqe_scaling}). Here, the former systematically overestimate the experimental cross section for almost all data points, while the latter, although describing the slopes fairly well, underestimate the peak cross section. The red lines mark the peak cross section values given by Eq.~(50) from Ref.~\cite{Kocevar:1967}, which reasonably match the experiment. The peak values given by Eq.~\eqref{eq:durand} are not shown here, since this approximation does not work well for those high values of $Q^{2}$.

To estimate more quantitatively the overall quality of the data description by the Bosted parameterization, a comparison of the corresponding integrals under the quasi-elastic peak was performed. For this purpose all distributions were integrated within the limits shown by the vertical lines in Figs.~\ref{fig:hanson_QE} and~\ref{fig:rock_QE}. To determine the positions of these limits, the quasi-elastic peaks in the experimental spectra were fit by Gaussians with polynomial background. Then the values $\mu-\sigma$ and $\mu+3\sigma$ were set as the left and right integration limits, respectively, with $\mu$ and $\sigma$ being the mean value and the standard deviation of the corresponding Gaussian function. The integration limits were chosen to be asymmetrical in order to minimize the inelastic background under the quasi-elastic peak. This procedure of obtaining the integration limits was used in order to achieve consistency among all plots, since the width of the quasi-elastic peak and its proximity to the inelastic part of the spectrum depend on the kinematics.

The results of the performed comparison are summarized in Tab.~\ref{tab:quasi_el_tab}, where the first six rows correspond to the first dataset from Ref.~\cite{Hanson:1973vf}, while the last two correspond to the second dataset from  Refs.~\cite{Rock:1991jy,Rock_SLAC}. The last four columns contain the values of the ratio of the experimental integral under the quasi-elastic peak ($\sigma_{exp}$) to that obtained from the Bosted parameterization, in which $F^{QE}$ is calculated by using the Paris wave function ($\sigma_{par}^{1}$) or given by Eq.~\eqref{eq:fqe_scaling} ($\sigma_{par}^{2}$). The index $norm$ indicates that the parameterization histograms were scaled in a way that their maxima were equal to the predictions of Eq.~\eqref{eq:durand} for the first dataset and to the predictions of Eq.~(50) from Ref.~\cite{Kocevar:1967} for the second dataset. The coloring of the table cells is related to the corresponding deviation of the obtained ratio from unity\footnote[6]{The ratio values given in Tab.~\ref{tab:quasi_el_tab} are slightly dependent on the positions of the integration limits, but the overall behavior of the data description quality reflected in the table is stable.}: the dark-green shade stands for deviations $\leq 5$\%, light-green for 5\%-10\%, and light-red for more than 10\%. 

As is seen from Tab.~\ref{tab:quasi_el_tab}, the Bosted parameterization with $F_{QE}$ calculated by the default method systematically overestimates the measured integral cross sections under the quasi-elastic peak, while with $F_{QE}$ calculated in the alternative way it systematically underestimates them. Beside this, when normalized to the values provided by the corresponding peak cross section approximations~\cite{Durand:1961zz,Kocevar:1967}, the Bosted parameterization better describes the integral cross sections for the majority of considered measurements.

%% file: text/concl/concl.tex
\newpage
\chapter{Conclusion}
\mbox{}\vspace{-\baselineskip}

A sophisticated testing of several parameterizations~\cite{Bosted_fit,Bosted:2007xd,Durand:1961zz,Kocevar:1967} existing for the deuteron quasi-elastic peak was performed via the comparison of the parameterized cross sections with two published sets of experimental cross sections~\cite{Hanson:1973vf,Rock:1991jy,Rock_SLAC}. The comparison was made in the wide $Q^{2}$ range from $\sim$0.3~GeV$^{2}$ to $\sim$4~GeV$^{2}$. This impartial examination allows to make the following conclusions.

\begin{itemize}

\item The Bosted parameterization~\cite{Bosted_fit,Bosted:2007xd} in its default implementation systematically overestimates the measured integral cross sections under the quasi-elastic peak. The overall data description quality gradually decreases from several percent to almost 20\% as $Q^2$ grows from 0.3~GeV$^{2}$ to 4~GeV$^{2}$.

\item The Bosted parameterization~\cite{Bosted_fit,Bosted:2007xd} with $F_{QE}$ calculated according to Eq.~\eqref{eq:fqe_scaling} systematically underestimates the measured integral cross sections under the quasi-elastic peak. The overall data description quality gradually increases from $\sim$15\% to a few percent as $Q^2$ grows from 0.3~GeV$^{2}$ to 4~GeV$^{2}$.

\item The normalization of the parametrized distributions to the values provided by the corresponding peak cross section approximations~\cite{Durand:1961zz,Kocevar:1967} gives some improvement in the description quality for the majority of considered measurements.

\end{itemize}

Thus, in the considered $Q^{2}$ range the Bosted parameterization of the deuteron quasi-elastic peak (in its default implementation) was found to give worse data description quality than that offered by the Bosted parameterization of the proton elastic peak (see App.~\ref{app_epelas}). This may serve as an indication that internal structure of the deuteron is not yet fully understood.

In addition to this examination, some helpful tools that may be of use for studying quasi-elastic and inclusive cross sections off deuterons are given in App.~\ref{app_tools}.

%% file: text/append/append.tex
\appendix

 \refstepcounter{chapter}
    \makeatletter
   \renewcommand{\theequation}{\thechapter.\@arabic\c@equation}
    \makeatother
\chapter*{Appendices}
\label{app}
\addcontentsline{toc}{chapter}{Appendices}

\renewcommand{\thesection}{A}
 \refstepcounter{section}
    \makeatletter
   \renewcommand{\theequation}{\thesection.\@arabic\c@equation}
    \makeatother
\section*{Appendix A: Formalism of the $ep$ scattering}
\label{app_formalism}
\addcontentsline{toc}{section}{A: Formalism of the $ep$ scattering}

\subsection*{Elastic $ep$ scattering}

The cross section for the elastic scattering of an electron off a nucleon~\cite{Halzen:1984mc,Close:1979bt,Povh:1995mua} can be described as:
\begin{equation}
\frac{d^{2}\sigma}{d\Omega d E'} = \left [ \frac{d\sigma}{d\Omega} \right ]^{*}_{Mott}  \left [ \frac{G_{E}^{2}(Q^{2}) + \tau G_{M}^{2}(Q^{2}) }{1+\tau} + 2\tau G_{M}^{2}(Q^{2}) \tan^{2}{\frac{\theta_{e'}}{2}} \right ] \delta\left ( \nu - \frac{Q^{2}}{2m_{N}} \right),\label{eq_p_elast1}
\end{equation}
where $E'$ and $\theta_{e'}$ are the energy and polar angle of the scattered electron in the laboratory frame, respectively, $\nu = E-E'$ is the photon energy with $E$ being the laboratory energy of the incoming electron, $Q^{2}$ the photon virtuality, $m_{N}$ the nucleon mass, and $\tau = \frac{|Q^{2}|}{4m_{N}^{2}}$.

In Eq.~\eqref{eq_p_elast1} the Mott cross section\footnote[7]{Following the notation of  Ref.~\cite{Povh:1995mua}, the asterisk superscript indicates that Mott cross section formula does not include the factor $E'/E$.} is defined according to Eq.~\eqref{eq:mott} and $G_{E}(Q^{2})$ and $G_{M}(Q^{2})$ are the electric and magnetic nucleon form factors (or Sachs form factors).

Upon integration over $d E'$, Eq.~\eqref{eq_p_elast1} takes the following form.\footnote[8]{The following relation was used $\int dx \delta\left ( f(x) \right ) = \left [ df/dx \right ]^{-1}$.}
\begin{equation}
\frac{d\sigma}{d\Omega} = \left [ \frac{d\sigma}{d\Omega} \right ]^{*}_{Mott} \left [1+\frac{2E}{m_{N}}\sin^{2} \frac{\theta_{e'}}{2} \right ]^{-1}   \left [ \frac{G_{E}^{2}(Q^{2}) + \tau G_{M}^{2}(Q^{2}) }{1+\tau} + 2\tau G_{M}^{2}(Q^{2}) \tan^{2}{\frac{\theta_{e'}}{2}} \right ].\label{eq_p_elast2}
\end{equation}

Taking into account the fact that according to the energy conservation
\begin{equation}
1+\frac{2E}{m_{N}}\sin^{2} \frac{\theta_{e'}}{2} = \frac{1}{E'}\left ( E' + \frac{Q^{2}}{2m_{N}} \right ) = \frac{E}{E'}, 
\end{equation}
one can obtain the more commonly used formula, i.e. 
\begin{equation}
\frac{d\sigma}{d\Omega} = \left [ \frac{d\sigma}{d\Omega} \right ]^{*}_{Mott} \left [ \frac{E'}{E} \right ]   \left [ \frac{G_{E}^{2}(Q^{2}) + \tau G_{M}^{2}(Q^{2}) }{1+\tau} + 2\tau G_{M}^{2}(Q^{2}) \tan^{2}{\frac{\theta_{e'}}{2}} \right ],\label{eq_p_elast3}
\end{equation}
where the factor $E'/E$ accounts for the recoil of the target nucleon.

Meanwhile, the elastic scattering cross section is often written in terms of the form factors $F_{1}(Q^{2})$ and $F_{2}(Q^{2})$ as
\begin{equation}
\frac{d\sigma}{d\Omega} = \left [ \frac{d\sigma}{d\Omega} \right ]^{*}_{Mott} \left [ \frac{E'}{E} \right ]   \left [ F^{2}_{1}(Q^{2}) + \chi^{2} \tau F^{2}_{2}(Q^{2})  + 2\tau \left [F_{1}(Q^{2}) + \chi F_{2}(Q^{2}) \right ]^{2} \tan^{2}{\frac{\theta_{e'}}{2}} \right ],\label{eq_p_elast3}
\end{equation}
where $\chi$ is the nucleon magnetic moment.

The relations between $G_{E}(Q^{2})$,  $G_{M}(Q^{2})$ and $F_{1}(Q^{2})$, $F_{2}(Q^{2})$ are the following.
\begin{equation}
\begin{aligned}
&G_{E}(Q^{2}) &=~& F_{1}(Q^{2}) - \chi \tau F_{2}(Q^{2})\\
&G_{M}(Q^{2}) &=~& F_{1}(Q^{2}) + \chi F_{2}(Q^{2})
\label{eq:el_f1_f2}
\end{aligned}
\end{equation}

\subsection*{Inelastic $ep$ scattering}

By analogy with the elastic scattering case, the cross section for the inelastic scattering of an electron off a nucleon~\cite{Halzen:1984mc,Close:1979bt,Povh:1995mua} can be described as:
\begin{equation}
\frac{d^{2}\sigma}{d\Omega d E'} = \left [\frac{\alpha}{Q^2}\right ]^{2} \left [\frac{E'}{E}\right ] L_{e}^{\mu \nu}W_{\mu \nu} =  \left [ \frac{d\sigma}{d\Omega} \right ]^{*}_{Mott}   \left [ W_{2}(Q^{2},\nu) + 2W_{1}(Q^{2},\nu)\tan^{2}{\frac{\theta_{e'}}{2}} \right ],\label{eq:p_inelast}
\end{equation}
where $L_{e}^{\mu \nu}W_{\mu \nu}$ is the convolution of the leptonic and hadronic tensors, while $W_{1}(Q^{2},\nu)$ and $W_{2}(Q^{2},\nu)$ are two dimensionful structure functions. They are usually replaced by the corresponding two dimensionless structure functions as
\begin{equation}
\begin{aligned}
&F_{1}(Q^{2},x) &=~& m_{N}W_{1}(Q^{2},\nu)~\textrm{and} \\
&F_{2}(Q^{2},x) &=~& \nu W_{2}(Q^{2}, \nu),
\label{eq:dimless_str_f}
\end{aligned}
\end{equation}
where $x=\frac{Q^{2}}{2 \nu m_{N}}$ is the Bjorken scaling variable. One should not confuse these structure functions with $F_{1}(Q^{2})$ and $F_{2}(Q^{2})$ elastic scattering form factors from Eqs.~\eqref{eq_p_elast3} and~\eqref{eq:el_f1_f2}.

Comparing Eq.~\eqref{eq:p_inelast} with Eq.~\eqref{eq_p_elast1} one can conclude that for the elastic scattering~\cite{Close:1979bt} 
\begin{equation}
\begin{aligned}
&W_{1}^{el}(Q^{2},\nu) &=~& \tau G_{M}^{2}(Q^{2})\delta\left ( \nu - \frac{Q^{2}}{2m_{N}} \right)~\textrm{and}  \\
&W_{2}^{el}(Q^{2},\nu) &=~& \frac{G_{E}^{2}(Q^{2}) + \tau G_{M}^{2}(Q^{2}) }{1+\tau} \delta\left ( \nu - \frac{Q^{2}}{2m_{N}} \right).
\label{eq:rel_str_fun}
\end{aligned}
\end{equation}

\newpage

\setcounter{table}{0}
\renewcommand{\thetable}{B.\arabic{table}}

\renewcommand{\thesection}{B}
 \refstepcounter{section}
    \makeatletter
   \renewcommand{\theequation}{\thesection.\@arabic\c@equation}
    \makeatother
\section*{B: Measured elastic $ep$ cross sections versus parameterized}
\label{app_epelas}
\addcontentsline{toc}{section}{B: Measured elastic $ep$ cross sections versus parameterized}

\begin{table}[htp]

\caption{\small Values of experimental $ep$ elastic cross sections ($\left [ \frac{d\sigma}{d\Omega} \right ]_{exp}$) provided with their total uncertainties $\varepsilon_{exp}$ and the corresponding values obtained from the Bosted parameterization ($\left [ \frac{d\sigma}{d\Omega} \right ]_{par}$). The last column contains their ratios. The coloring of the table cells is related to the corresponding deviation of the obtained ratio from unity: the dark-green shade stands for deviations $\leq 5$\%, light-green for 5\%-10\%, and light-red shows deviations of more than 10\%. Experimental values$^\dag$~were taken from Refs.~\cite{Goitein:1970pz,Sill:1992qw,Christy:2004rc} and were picked up in a way that they cover relatively evenly the $Q^{2}$ range from $\sim$0.3~GeV$^{2}$ to $\sim$5~GeV$^{2}$.}\label{tab:epelast}
\centering
\begin{tabular}{
   !{\vrule width 1pt}
  c!{\vrule width 1pt}
  c!{\vrule width 1pt}
  c!{\vrule width 1pt}
  c!{\vrule width 1pt}
  c!{\vrule width 1pt}
  c!{\vrule width 1pt}
  c!{\vrule width 1pt}
  }
\toprule[2pt]
\makecell{Exp.\\ Ref.}& \makecell{$E_{beam}$\\ (GeV)} &\makecell{$Q^{2}$\\ (GeV$^2$)} & \makecell{$\left [ \frac{d\sigma}{d\Omega} \right ]_{exp}$\\ (nb/sr)}&\makecell{$\varepsilon_{exp}$ (\%)} & \makecell{$\left [ \frac{d\sigma}{d\Omega} \right ]_{par}$\\ (nb/sr)}& $\left [ \frac{d\sigma}{d\Omega} \right ]_{exp}/\left [ \frac{d\sigma}{d\Omega} \right ]_{par}$  \\ \hline
\multirow{6}{*}{\cite{Goitein:1970pz}}& 1.578  & 0.2725 &7.675E2  & 2.1  &7.808E2   & \cellcolor{green!35}0.98 \\ \hhline{|~|------|}
& 3.440  & 1.168  &2.304E1  & 2.9  &2.313E1   & \cellcolor{green!35}1.00 \\ \hhline{|~|------|}
& 4.308  & 1.752  &5.792E0  & 3.3  &6.088E0   & \cellcolor{green!35}0.95 \\ \hhline{|~|------|}
& 5.500  & 2.725  &1.209E0  & 5.2  &1.126E0   & \cellcolor{green!20}1.07 \\ \hhline{|~|------|}
& 6.000  & 3.504  &3.64E-1  & 4.6  &3.280E-1   & \cellcolor{red!20}1.11 \\ \hhline{|~|------|}
\multirow{6}{*}{\cite{Sill:1992qw}}& 6.000  & 4.478  &7.25E-2  & 5.6  &6.606E-2   & \cellcolor{green!20}1.10 \\ \midrule[2pt]
& 5.464  & 2.862  &8.02E-1  & 3.8  &8.361E-1   & \cellcolor{green!35}0.96 \\ \hhline{|~|------|}
& 5.464  & 3.621  &1.93E-1  & 4.2  &1.974E-1   & \cellcolor{green!35}0.98  \\ \hhline{|~|------|}
\multirow{8}{*}{\cite{Christy:2004rc}}& 5.499  & 5.017  &2.04E-2  & 3.7  &2.076E-2   & \cellcolor{green!35}0.98 \\ \midrule[2pt]
& 1.148  & 0.62   &1.734E1  & 1.9     &1.735E1   & \cellcolor{green!35}1.00 \\\hhline{|~|------|}
& 2.235  & 1.6348 &1.184E0  & 2.0     &1.180E0    & \cellcolor{green!35}1.00 \\ \hhline{|~|------|}
& 3.114  & 2.6205 &2.125E-1 & 2.0     &2.153E-1  &\cellcolor{green!35} 0.99 \\ \hhline{|~|------|}
& 4.104  & 3.7981 &4.919E-2 & 2.0     &4.865E-2  & \cellcolor{green!35}1.01  \\ \hhline{|~|------|}
& 4.413  & 4.7957 &1.098E-2 & 2.7     &1.082E-2  & \cellcolor{green!35}1.02  \\\hhline{|~|------|}
& 5.494  & 5.3699 &1.267E-2 & 2.5     &1.216E-2  & \cellcolor{green!35}1.04  \\ \bottomrule[2pt]
\end{tabular}
 \end{table}

$^\dag$\footnotesize {\centering Note that studies in Refs.~\cite{Goitein:1970pz,Sill:1992qw,Christy:2004rc} contain more measurements than given in the table, which is intended to reflect the overall behavior of the data description and hence shows just a small sample of typical examples.}
\normalsize 
\newpage

\renewcommand{\thesection}{C}
 \refstepcounter{section}
    \makeatletter
   \renewcommand{\theequation}{\thesection.\@arabic\c@equation}
    \makeatother
\section*{C: Some related tools}
\label{app_tools}
\addcontentsline{toc}{section}{C: Some related tools}

The following tools may be of use for those interested in studying quasi-elastic and inclusive deuteron spectra.

\begin{itemize}

\item The set of various fits by Peter Bosted, which includes but is not limited to the parameterizations for quasi-elastic and inelastic structure functions for nuclei, is located on Peter Bosted's web page at \url{https://userweb.jlab.org/~bosted/fits.html}. It contains FORTRAN subroutines provided with a C++ Wrapper package.

\item The FORTRAN subroutine {\it elas} that returns the elastic cross section of electron scattering off free protons as a function of the electron scattering angle is available at \url{https://github.com/skorodumina/deuteron\_quasi\_elastic/blob/master/PBosted\_model/F1F209.f}. This subroutine employs a Bosted parameterization of nucleons electromagnetic form factors according to Ref.~\cite{Bosted:1994tm}.

\item The FORTRAN subroutine {\it elasrad} that returns radiated elastic cross section of electron scattering off free protons is located at \url{https://github.com/skorodumina/deuteron\_quasi\_elastic/blob/master/PBosted\_model/F1F209.f}. This subroutine employs radiative effects according to Ref.~\cite{Mo:1968cg}.

\item The description of QUEGG, which is a Monte Carlo event generator for quasi-elastic scattering on deuterium, is given in Ref.~\cite{QUEGG} together with the link to the source code and the running instructions.

\item A slightly modified version of the QUEGG source code that provides BOS output is available at \url{https://github.com/skorodumina/deuteron\_quasi\_elastic/tree/master/QUEEG}.

\item An alternative event generator for both quasi-elastic and inclusive radiated spectra off the deuteron exists, which is based on data from Ref.~\cite{Osipenko_f2note,Osipenko:2005gt}. To use this event generator one should contact directly Dr. Mikhail Osipenko.

 \end{itemize}

\newpage